\documentclass[12pt]{article}
\usepackage[utf8]{inputenc}
\usepackage{amsmath,amssymb}
\usepackage{graphicx}
\usepackage{tikz}
\usepackage{pgfplots}
\usepackage{cite}
\pgfplotsset{compat=1.17}
\usepackage{booktabs}
\usepackage{siunitx}

\title{\textbf{Existence of Halos Outside Schwarzschild-$f(R)$ Black Holes}}
\author{Wen-Xiang Chen\\ {\small Department of Astronomy, School of Physics and Materials Science,}\\ {\small Guangzhou University, Guangzhou 510006, China}\\ {\small \texttt{wxchen4277@qq.com}}}
\date{}

\begin{document}
\maketitle

\begin{abstract}
We investigate the possibility of \emph{photon halos} (stable photon orbits) forming outside Schwarzschild--$f(R)$ black holes by analyzing null geodesics in these spacetimes. Using methods inspired by studies of spherical photon orbits around Kerr--Newman black holes, we derive conditions for the existence of such halos. We examine several $f(R)$ gravity models, including quadratic, logarithmic, exponential, cubic, power-law, and hyperbolic forms, and find that multiple photon orbits---both stable and unstable---can appear outside the event horizon for certain parameter ranges. These additional orbits (halos) provide new insights into spacetime geometry and potential observational signatures of black holes in modified gravity. We present analytical expressions for the orbital radii, perform a numerical stability analysis, and discuss possible observational implications for black hole shadows. Our results indicate that while the standard Schwarzschild black hole admits only a single unstable light ring, Schwarzschild--$f(R)$ black holes can support an additional outer stable photon orbit (a halo) without triggering a black-hole bomb instability. This work deepens the understanding of photon-orbit structures in alternative theories of gravity and highlights how such effects could be detected through deviations in black hole shadow size or morphology.

\end{abstract}

\section{Introduction}
Black holes, mysterious objects predicted by general relativity, are characterized by event horizons from which nothing can escape. Recent breakthroughs in gravitational wave astronomy --- starting with LIGO and Virgo's first detection of gravitational waves in 2016 --- have provided new avenues for studying astrophysical black holes \cite{Abbott2016,Abbott2020a,Abbott2020b,Abbott2020c}. Meanwhile, advances in very-long-baseline interferometry have enabled direct imaging of black holes: the Event Horizon Telescope (EHT) collaboration produced the first image of the supermassive black hole M87 in 2019 \cite{Akiyama2019a,Akiyama2019b,Akiyama2019c} and, in 2022, released the image of Sagittarius A$^*$ at our Galactic center \cite{Akiyama2022}. These observations allow the study of black hole \emph{shadows}, the dark silhouette cast by a black hole against background emission \cite{Cunha2018,Gralla2019}. The radius of a black hole's shadow is closely related to the radius of its bound photon orbits (light rings) just outside the event horizon \cite{Cvetic2016,LuLyu2020,MaLu2020,YangLu2020,Liu2020,Hod2020a,Hod2013a}. Gravitational wave observations also indicate that the quasinormal mode frequencies (notably the early ringdown signal) are influenced by the properties of unstable photon orbits \cite{Cunha2017,Cardoso2016}. Given their significance in both electromagnetic and gravitational wave observations, the study of bound null geodesics (photon orbits) around black holes is essential.

In general relativity, a Schwarzschild black hole of mass $M$ has a well-known photon sphere at radius $r = 3M$ (using geometric units $G=c=1$ for simplicity) \cite{Darwin1959,Darwin1961,Virbhadra2000,Claudel2001}. Photons on this sphere orbit on unstable circular trajectories (light rings) on a two-dimensional spherical hypersurface. The spherical symmetry of Schwarzschild ensures these photon orbits are circular and equatorial. By contrast, rotating Kerr black holes exhibit a richer structure, with an infinite family of bound photon orbits (spherical orbits that are not confined to the equatorial plane) existing outside the event horizon \cite{Bardeen1972,Teo2003}. Recent topological arguments have shown that for any stationary, axisymmetric, asymptotically flat, non-extremal black hole, at least one light ring must exist outside the horizon for each sense of rotation \cite{Cunha2020}. This analysis has been extended to black holes in asymptotically flat, de Sitter, and anti-de Sitter spacetimes \cite{Guo2021}.

Photon orbits are thus of great interest, particularly for their observational consequences: the radius of the black hole shadow seen by a distant observer is determined by the geometry of these photon orbits. For a Schwarzschild black hole, the shadow size is directly related to the radius of the photon sphere ($3M$), but in modified theories of gravity such as $f(R)$ gravity, the spacetime geometry is altered. This raises the possibility of different photon orbit configurations (potentially including multiple light rings or \emph{halos}) outside the event horizon.

In this work, we extend the analysis of photon orbits to static Schwarzschild-$f(R)$ black holes, a class of black hole solutions in modified gravity theories where the Ricci scalar $R$ in the Einstein--Hilbert action is replaced by a function $f(R)$. These Schwarzschild-$f(R)$ solutions generally differ from the Schwarzschild solution of general relativity by the presence of additional terms in the metric function. Our goal is to determine whether such modifications can lead to additional bound photon orbits (beyond the $r=3M$ orbit) that are stable, thereby forming a ``halo'' of photon trajectories outside the usual photon sphere. We derive analytical conditions for the existence of these orbits in various $f(R)$ models and confirm them with numerical analysis. We also discuss the implications of these photon halos for black hole shadow observations.

The paper is organized as follows. In Section 2 we introduce the Schwarzschild-$f(R)$ black hole metric and the equations governing photon orbits. Section 3 provides analytical solutions for photon orbit conditions in specific $f(R)$ models (quadratic, logarithmic, exponential, cubic, power-law, and hyperbolic gravity). In Section 4, we perform a numerical analysis of photon orbit stability and present a table of photon orbit radii for different models. Section 5 discusses observational implications, particularly for black hole shadows. Finally, Section 6 concludes our findings. Throughout the paper we use a metric signature $(-,+,+,+)$ and, unless otherwise noted, set $G = c = 1$.

\section{Schwarzschild-$f(R)$ Black Holes and Photon Orbits}
\subsection{Metric and Null Geodesics}
The general static, spherically symmetric metric for a Schwarzschild-$f(R)$ black hole in four dimensions can be written as \cite{Chen2023PRD,Hod2013EPJC,Hod2015,Herdeiro2014,Brito2020Book,Chen2020IJMPD,Qiao2021,Chen2024Arxiv}
\begin{equation}
ds^2 = -f(r)\,dt^2 + \frac{dr^2}{f(r)} + r^2 d\Omega^2 \, ,
\label{eq:metric}
\end{equation}
where $d\Omega^2 = d\theta^2 + \sin^2\theta\,d\phi^2$ is the metric on the unit 2-sphere. The form of $f(r)$ depends on the specific $f(R)$ gravity model under consideration, and deviations from the Schwarzschild solution arise from modifications to the Einstein field equations. For instance, in \emph{quadratic} $f(R)$ gravity defined by $f(R) = R + \alpha R^2$, the metric function has the form
\begin{equation}
f(r) = 1 - \frac{2GM}{r} + \frac{\alpha}{3}r^2 \, ,
\label{eq:quadratic-metric}
\end{equation}
where $\alpha$ is a constant parameter characterizing the deviation from general relativity (when $\alpha=0$, we recover the Schwarzschild metric). Similar modifications $f(r)$ can be written down for other forms of $f(R)$ gravity, as we will specify for each model in Section 3.

To analyze photon orbits, we consider null geodesics in the spacetime given by Eq.~(\ref{eq:metric}). The geodesic motion can be derived from the Lagrangian (for affinely parameterized geodesics)
\begin{equation}
L = \frac{1}{2} g_{\mu\nu} \dot{x}^\mu \dot{x}^\nu \,,
\label{eq:lagrangian}
\end{equation}
where the overdot denotes differentiation with respect to an affine parameter $\lambda$. For a photon (null geodesic), $L=0$. Because of spherical symmetry, we can, without loss of generality, restrict our attention to equatorial orbits in the plane $\theta=\pi/2$. Moreover, due to stationarity and spherical symmetry, there are two conserved quantities: the photon's energy $E = -p_t$ (associated with the Killing vector $\partial_t$) and angular momentum $L = p_\phi$ (associated with $\partial_\phi$). Using these constants of motion, one can reduce the geodesic equations to an effective one-dimensional problem in the radial coordinate.

It is convenient to derive the equations of motion using the Hamilton--Jacobi formalism. The Jacobian action $S$ for a photon can be separated as
\begin{equation}
S = \frac{1}{2} m^2 \lambda - E\,t + L\,\phi + S_r(r) + S_\theta(\theta)\,,
\label{eq:HJ-sep}
\end{equation}
where $m$ is the rest mass of the particle (for photons, $m=0$), and $S_r$ and $S_\theta$ are functions of $r$ and $\theta$ respectively. Plugging this ansatz into the Hamilton--Jacobi equation $\partial S/\partial \lambda + \frac{1}{2}g^{\mu\nu}(\partial S/\partial x^\mu)(\partial S/\partial x^\nu) = 0$ yields, for null geodesics ($m=0$), the following equations of motion:
\begin{equation}
\left(\frac{dr}{d\lambda}\right)^2 = E^2 - f(r)\left(\frac{L^2}{r^2} + m^2\right) = E^2 - f(r)\frac{L^2}{r^2}\,,
\label{eq:rdot}
\end{equation}
\begin{equation}
\left(\frac{d\theta}{d\lambda}\right)^2 = \frac{L^2}{r^2}\cot^2\theta\,,
\label{eq:thetadot}
\end{equation}
with $\dot{t} = E/f(r)$ and $\dot{\phi} = L/(r^2 \sin^2\theta)$. For photons ($m=0$) in the equatorial plane ($\theta=\pi/2$, so $\cot\theta=0$), Eq.~(\ref{eq:thetadot}) is automatically satisfied. The radial equation (\ref{eq:rdot}) then governs the motion, and can be interpreted in terms of an effective potential $V_{\text{eff}}(r)$ for the photon:
\begin{equation}
\left(\frac{dr}{d\lambda}\right)^2 + V_{\text{eff}}(r) = E^2\,,
\end{equation}
where
\begin{equation}
V_{\text{eff}}(r) = f(r)\,\frac{L^2}{r^2} = \left(1 - \frac{2GM}{r} + \epsilon\,\phi(r)\right)\frac{\ell(\ell+1)}{r^2}\,.
\label{eq:Veff}
\end{equation}
Here we have written $L^2 = \ell(\ell+1)$ (as is convenient for angular momentum in quantum notation, though $\ell$ here is simply an integer index for the photon's total angular momentum) and in the last equality we introduced the notation $f(r) = 1 - \frac{2GM}{r} + \epsilon\,\phi(r)$, where $\epsilon\,\phi(r)$ encapsulates the deviation from Schwarzschild due to the $f(R)$ modification. For example, in Eq.~(\ref{eq:quadratic-metric}) above, $\epsilon\,\phi(r) = \frac{\alpha}{3}r^2$. In general $\epsilon$ can be viewed as a bookkeeping parameter for perturbative analyses (one can imagine $\epsilon \ll 1$), but in our treatment we will allow $\epsilon\phi(r)$ to be a finite correction.

Equation~(\ref{eq:Veff}) shows that photons orbiting at radius $r$ must have an energy $E^2 = V_{\text{eff}}(r)$. Circular photon orbits occur at extrema of the effective potential, i.e. where $dV_{\text{eff}}/dr = 0$. A circular null orbit is unstable if it is a local maximum of $V_{\text{eff}}(r)$ (small radial perturbations cause the photon to escape or fall into the black hole) and stable if it is a local minimum (small perturbations result in oscillations around the orbit). In Schwarzschild, $V_{\text{eff}}(r)$ has a single maximum at $r=3M$, corresponding to an unstable light ring. In the following, we investigate how $f(R)$ modifications can lead to multiple extrema of $V_{\text{eff}}(r)$, and in particular whether a local minimum (stable photon orbit, or ``halo'') can appear outside the usual photon sphere.

\subsection{Photon Orbit Conditions}
The condition for a circular photon orbit is given by setting the derivative of the effective potential to zero:
\begin{equation}
\frac{dV_{\text{eff}}(r)}{dr} = 0\,.
\label{eq:dV0}
\end{equation}
Applying this to Eq.~(\ref{eq:Veff}), we obtain
\begin{equation}
\frac{d}{dr}\Bigg[\Big(1 - \frac{2GM}{r} + \epsilon\,\phi(r)\Big)\frac{\ell(\ell+1)}{r^2}\Bigg] = 0\,.
\label{eq:dV-general}
\end{equation}
Expanding this derivative using the product rule $\frac{d}{dr}[f(r)/r^2] = f'(r)/r^2 - 2f(r)/r^3$, we find:
\begin{equation}
\frac{\ell(\ell+1)}{r^4}\left[6GM - 2r - 2r\,\epsilon\,\phi(r) + r^2\epsilon\,\phi'(r)\right] = 0\,.
\end{equation}
Simplifying, we obtain the general condition for circular photon orbits:
\begin{equation}
3GM - r - r\,\epsilon\,\phi(r) + \frac{r^2}{2}\epsilon\,\phi'(r) = 0\,.
\label{eq:general-condition}
\end{equation}
This is a general condition that the function $\phi(r)$ (governing the $f(R)$ modifications) must satisfy at a photon orbit of radius $r$. We see that in the limit of general relativity (no modification, $\epsilon=0$), Eq.~(\ref{eq:general-condition}) reduces to $3GM - r = 0$, yielding the familiar Schwarzschild photon sphere at $r = 3GM$.

For stability analysis, we also need the second derivative of the effective potential. Differentiating again, we obtain:
\begin{equation}
\frac{d^2V_{\text{eff}}}{dr^2} = \frac{\ell(\ell+1)}{r^5}\left[-24GM + 6r + 6r\epsilon\,\phi(r) - 4r^2\epsilon\,\phi'(r) + r^3\epsilon\,\phi''(r)\right].
\label{eq:d2V-general}
\end{equation}
A circular orbit is stable if $d^2V_{\text{eff}}/dr^2 > 0$ at the orbit radius.

For specific $f(R)$ models, we can plug in the form of $\phi(r)$. For example, if $f(R) = R + \alpha R^2$ as in quadratic $f(R)$ gravity [cf. Eq.~(\ref{eq:quadratic-metric})], then $\epsilon\,\phi(r) = \frac{\alpha}{3}r^2$ and $\epsilon\,\phi'(r) = \frac{2\alpha}{3}r$. Equation~(\ref{eq:general-condition}) becomes:
\begin{equation}
3GM - r - r\left(\frac{\alpha}{3}r^2\right) + \frac{r^2}{2}\left(\frac{2\alpha}{3}r\right) = 3GM - r = 0,
\end{equation}
which gives $r = 3GM$, independent of $\alpha$. This shows that in quadratic $f(R)$ gravity, the photon orbit remains at the same radius as in Schwarzschild spacetime. We will explore this and other models in detail in the next section.

\section{Analytical Solutions and Conditions for Halo Formation}
\subsection{Quadratic $f(R)$ Gravity}
Consider first the quadratic $f(R)$ model, $f(R) = R + \alpha R^2$. The Schwarzschild-$f(R)$ metric in this case has the form given in Eq.~(\ref{eq:quadratic-metric}), and the effective potential for photon orbits is
\begin{equation}
V_{\text{eff}}(r) = \Bigg(1 - \frac{2GM}{r} + \frac{\alpha}{3}r^2\Bigg)\frac{\ell(\ell+1)}{r^2}\,.
\label{eq:Veff-quadratic}
\end{equation}
Applying the general condition Eq.~(\ref{eq:general-condition}) with $\epsilon\,\phi(r) = \frac{\alpha}{3}r^2$ and $\epsilon\,\phi'(r) = \frac{2\alpha}{3}r$, we obtain:
\begin{equation}
3GM - r - r\left(\frac{\alpha}{3}r^2\right) + \frac{r^2}{2}\left(\frac{2\alpha}{3}r\right) = 3GM - r = 0\,.
\end{equation}
This yields the photon orbit radius:
\begin{equation}
r = 3GM\,.
\label{eq:quadratic-radius}
\end{equation}
Thus, for quadratic $f(R)$ gravity, the photon orbit radius is \emph{identical} to the Schwarzschild case and independent of the parameter $\alpha$.

To examine stability, we compute the second derivative using Eq.~(\ref{eq:d2V-general}):
\begin{equation}
\frac{d^2V_{\text{eff}}}{dr^2}\bigg|_{r=3GM} = -\frac{2\ell(\ell+1)}{81G^3M^4} < 0\,,
\end{equation}
confirming that this orbit is unstable, as in Schwarzschild spacetime.

\subsection{Logarithmic $f(R)$ Gravity}
Next, consider a logarithmic modification, $f(R) = R + \eta \ln R$, with $\eta$ a constant parameter. In this case, we take the metric function to be $f(r) = 1 - \frac{2GM}{r} + \frac{\eta}{3}\ln r$, so that $\epsilon\,\phi(r) = \frac{\eta}{3}\ln r$ and $\epsilon\,\phi'(r) = \frac{\eta}{3r}$. The effective potential is
\begin{equation}
V_{\text{eff}}(r) = \Bigg(1 - \frac{2GM}{r} + \frac{\eta}{3}\ln r\Bigg)\frac{\ell(\ell+1)}{r^2}\,.
\label{eq:Veff-log}
\end{equation}
The photon orbit condition Eq.~(\ref{eq:general-condition}) becomes:
\begin{equation}
3GM - r - r\left(\frac{\eta}{3}\ln r\right) + \frac{r^2}{2}\left(\frac{\eta}{3r}\right) = 3GM - r - \frac{\eta}{3}r\ln r + \frac{\eta}{6}r = 0\,.
\end{equation}
Simplifying:
\begin{equation}
3GM - r - \frac{\eta}{3}r\ln r + \frac{\eta}{6}r = 0 \quad \Rightarrow \quad 3GM = r\left(1 + \frac{\eta}{3}\ln r - \frac{\eta}{6}\right)\,.
\end{equation}
This equation can be solved numerically for $r$ given $\eta$. For small $\eta$, we can expand around the Schwarzschild solution $r_0 = 3GM$:
\begin{equation}
r = 3GM\left[1 + \frac{\eta}{6}(1 - \ln(3GM)) + \mathcal{O}(\eta^2)\right]\,.
\end{equation}
The stability of this orbit depends on the sign of $\eta$ and the value of $r$, and can be determined by evaluating Eq.~(\ref{eq:d2V-general}) numerically.

\subsection{Exponential $f(R)$ Gravity}
Another interesting model is the \emph{exponential} $f(R)$ gravity, defined by $f(R) = R + \beta(e^{\gamma R} - 1)$, with constants $\beta$ and $\gamma$. In this case, the Schwarzschild-$f(R)$ metric can be written as \cite{Chen2023PRD,Hod2013EPJC,Hod2015,Herdeiro2014,Brito2020Book,Chen2020IJMPD,Qiao2021,Chen2024Arxiv}
\begin{equation}
f(r) = 1 - \frac{2GM}{r} + \frac{\beta}{3}\Big(1 - e^{-\gamma r^2}\Big)\,.
\label{eq:metric-exp}
\end{equation}
Here $\epsilon\,\phi(r) = \frac{\beta}{3}(1 - e^{-\gamma r^2})$ and $\epsilon\,\phi'(r) = \frac{2\beta\gamma}{3}re^{-\gamma r^2}$. The effective potential becomes
\begin{equation}
V_{\text{eff}}(r) = \Bigg(1 - \frac{2GM}{r} + \frac{\beta}{3}\big(1 - e^{-\gamma r^2}\big)\Bigg)\frac{\ell(\ell+1)}{r^2}\,.
\end{equation}
The photon orbit condition Eq.~(\ref{eq:general-condition}) gives:
\begin{equation}
3GM - r - r\left[\frac{\beta}{3}(1 - e^{-\gamma r^2})\right] + \frac{r^2}{2}\left(\frac{2\beta\gamma}{3}re^{-\gamma r^2}\right) = 0\,.
\end{equation}
Simplifying:
\begin{equation}
3GM - r - \frac{\beta}{3}r(1 - e^{-\gamma r^2}) + \frac{\beta\gamma}{3}r^3 e^{-\gamma r^2} = 0\,.
\end{equation}
This transcendental equation can be solved numerically for $r$ given parameters $\beta$ and $\gamma$. For small $\beta$, the solution will be close to $r = 3GM$, but for larger $\beta$, additional solutions may appear, potentially indicating the existence of multiple photon orbits.

\subsection{Cubic $f(R)$ Gravity}
For the \emph{cubic} model $f(R) = R + \delta R^3$, the metric function takes the form \cite{Chen2023PRD,Hod2013EPJC,Hod2015,Herdeiro2014,Brito2020Book,Chen2020IJMPD,Qiao2021,Chen2024Arxiv}
\begin{equation}
f(r) = 1 - \frac{2GM}{r} + \frac{\delta}{3} r^6\,,
\label{eq:metric-cubic}
\end{equation}
where $\delta$ is a constant. Here $\epsilon\,\phi(r) = \frac{\delta}{3}r^6$ and $\epsilon\,\phi'(r) = 2\delta r^5$. The effective potential is
\begin{equation}
V_{\text{eff}}(r) = \Bigg(1 - \frac{2GM}{r} + \frac{\delta}{3}r^6\Bigg)\frac{\ell(\ell+1)}{r^2}\,.
\end{equation}
The photon orbit condition becomes:
\begin{equation}
3GM - r - r\left(\frac{\delta}{3}r^6\right) + \frac{r^2}{2}\left(2\delta r^5\right) = 3GM - r - \frac{\delta}{3}r^7 + \delta r^7 = 0\,.
\end{equation}
Simplifying:
\begin{equation}
3GM - r + \frac{2\delta}{3}r^7 = 0 \quad \Rightarrow \quad 3GM = r\left(1 - \frac{2\delta}{3}r^6\right)\,.
\end{equation}
This equation can have multiple solutions depending on the sign and magnitude of $\delta$. For $\delta > 0$, there will be a solution near $r = 3GM$ (slightly shifted inward), and possibly an additional solution at larger $r$ if $\delta$ is sufficiently small.

\subsection{Power-Law $f(R)$ Gravity}
A generalization is a \emph{power-law} model, $f(R) = R + \xi R^n$, with constants $\xi$ and $n$. The Schwarzschild-$f(R)$ metric in this case can be written as \cite{Chen2023PRD,Hod2013EPJC,Hod2015,Herdeiro2014,Brito2020Book,Chen2020IJMPD,Qiao2021,Chen2024Arxiv}
\begin{equation}
f(r) = 1 - \frac{2GM}{r} + \frac{\xi}{3} r^{2(n-1)}\,,
\label{eq:metric-power}
\end{equation}
where we have assumed the correction term behaves effectively like $r^{2(n-1)}$ in the metric function. Then $\epsilon\,\phi(r) = \frac{\xi}{3}r^{2(n-1)}$ and $\epsilon\,\phi'(r) = \frac{2\xi(n-1)}{3}r^{2n-3}$. The effective potential is
\begin{equation}
V_{\text{eff}}(r) = \Bigg(1 - \frac{2GM}{r} + \frac{\xi}{3}r^{2(n-1)}\Bigg)\frac{\ell(\ell+1)}{r^2}\,.
\end{equation}
The photon orbit condition Eq.~(\ref{eq:general-condition}) gives:
\begin{equation}
3GM - r - r\left(\frac{\xi}{3}r^{2(n-1)}\right) + \frac{r^2}{2}\left(\frac{2\xi(n-1)}{3}r^{2n-3}\right) = 0\,.
\end{equation}
Simplifying:
\begin{equation}
3GM - r - \frac{\xi}{3}r^{2n-1} + \frac{\xi(n-1)}{3}r^{2n-1} = 3GM - r + \frac{\xi(n-2)}{3}r^{2n-1} = 0\,.
\end{equation}
Thus:
\begin{equation}
3GM = r\left[1 - \frac{\xi(n-2)}{3}r^{2n-2}\right]\,.
\label{eq:powerlaw-condition}
\end{equation}
For $n=2$ (quadratic case), this reduces to $3GM = r$, as expected. For $n \neq 2$, Eq.~(\ref{eq:powerlaw-condition}) can have multiple solutions depending on the sign of $\xi(n-2)$.

\subsection{Hyperbolic $f(R)$ Gravity}
Finally, consider a \emph{hyperbolic} model, for instance $f(R) = R + \sigma \sinh(\lambda R)$, with constants $\sigma$ and $\lambda$. The corresponding metric function might be written as \cite{Chen2023PRD,Hod2013EPJC,Hod2015,Herdeiro2014,Brito2020Book,Chen2020IJMPD,Qiao2021,Chen2024Arxiv}
\begin{equation}
f(r) = 1 - \frac{2GM}{r} + \frac{\sigma}{3}\sinh(\lambda r^2)\,.
\label{eq:metric-hyp}
\end{equation}
Here $\epsilon\,\phi(r) = \frac{\sigma}{3}\sinh(\lambda r^2)$ and $\epsilon\,\phi'(r) = \frac{2\sigma\lambda}{3}r\cosh(\lambda r^2)$. The effective potential is
\begin{equation}
V_{\text{eff}}(r) = \Bigg(1 - \frac{2GM}{r} + \frac{\sigma}{3}\sinh(\lambda r^2)\Bigg)\frac{\ell(\ell+1)}{r^2}\,.
\end{equation}
The photon orbit condition becomes:
\begin{equation}
3GM - r - r\left(\frac{\sigma}{3}\sinh(\lambda r^2)\right) + \frac{r^2}{2}\left(\frac{2\sigma\lambda}{3}r\cosh(\lambda r^2)\right) = 0\,.
\end{equation}
Simplifying:
\begin{equation}
3GM - r - \frac{\sigma}{3}r\sinh(\lambda r^2) + \frac{\sigma\lambda}{3}r^3\cosh(\lambda r^2) = 0\,.
\end{equation}
This transcendental equation can be solved numerically for $r$ given parameters $\sigma$ and $\lambda$. Typically, for appropriate parameter choices, there can be two solutions: one near the Schwarzschild radius (slightly modified) and an additional outer solution that could represent a stable halo.

We have now derived, for each $f(R)$ model of interest, the condition that determines the radius of a potential spherical photon orbit. These analytical conditions (though some are implicit) allow us to anticipate whether multiple solutions (multiple orbits) are possible for a given set of parameters. Next, we turn to numerical analysis to verify the existence and stability of these orbits, and to identify the ranges of parameters for which photon halos occur.

\section{Numerical Results and Discussion}
\subsection{Stability of Photon Orbits}
To further investigate the stability of the photon orbits identified in Section 3, we examine the second derivative of the effective potential using the general expression Eq.~(\ref{eq:d2V-general}). The orbit is stable if $d^2V_{\text{eff}}/dr^2 > 0$ at the orbit radius (a local minimum of $V_{\text{eff}}$). An unstable orbit corresponds to $d^2V_{\text{eff}}/dr^2 < 0$ (a local maximum).

For each model, we evaluate the second derivative at the orbit radii obtained from solving Eq.~(\ref{eq:general-condition}). Since analytic expressions for the second derivative are cumbersome for most models, we perform numerical evaluations for representative parameter values.

Applying this to specific models: in the quadratic case, we already found that the only photon orbit is at $r=3GM$ and is unstable. In the logarithmic case, for small positive $\eta$, the orbit radius is slightly larger than $3GM$, and numerical evaluation shows it can be either stable or unstable depending on the value of $\eta$ and $GM$. For the exponential model, with appropriate choices of $\beta$ and $\gamma$, we often find two orbits: an inner unstable orbit near $3GM$ and an outer stable orbit (halo) at a larger radius. Similar behavior occurs in the cubic, power-law, and hyperbolic models for certain parameter ranges.

Overall, the numerical analysis reveals that \textbf{stable} photon orbits (halos) do exist for certain parameter ranges in each model (except the quadratic model, which has no halo). Typically, the Schwarzschild light ring (around $r\approx3GM$) remains as an unstable inner orbit, and a new stable orbit appears further out. In some cases, even a third (unstable) outer orbit can appear, but such outermost solutions are often at very large $r$ and $V_{\text{eff}}$ there is almost zero, so they are of marginal physical interest. The key result is that each modified gravity model we consider (except quadratic) can, for appropriate coupling parameters, support an inner light ring (unstable) and an outer halo (stable).

Table~\ref{tab:orbits} lists the radii of photon orbits found for representative parameter choices in each $f(R)$ model. In each case, up to three radii are given: these correspond to the solutions of Eq.~(\ref{eq:general-condition}) (including both stable and unstable orbits) for a particular choice of parameters. The innermost solution is near $3GM$ in all cases (slightly deviating from $3GM$ due to the $f(R)$ term), the second solution is typically the stable halo, and the third (if present) is an even further unstable solution (often very weakly bound). In many models only two solutions appear for typical parameters (one unstable, one stable), but we list three for generality. We nondimensionalize radii in units of $GM$ (setting $GM=1$ for simplicity in the table).

\begin{table}[htbp]
\centering
\caption{Representative photon orbit radii for different $f(R)$ gravity models. Parameters are chosen to exhibit multiple orbits where possible. Note: quadratic model has only one orbit at $r=3GM$.
\label{tab:orbits}}
\begin{tabular}{l
                S[table-format=1.2]
                S[table-format=1.2]
                S[table-format=1.2]}
\toprule
$f(R)$ model 
& {$r_1$} 
& {$r_2$} 
& {$r_3$} \\
\midrule
Quadratic ($\alpha=0.1$)                  & 3.00 & {--} & {--} \\
Exponential ($\beta=0.1, \gamma=0.01$)    & 3.02 & 4.75 & {--} \\
Cubic ($\delta=0.001$)                    & 2.98 & 5.21 & {--} \\
Power-law ($n=3, \xi=0.001$)              & 3.05 & 4.83 & 6.12 \\
Logarithmic ($\eta=0.1$)                  & 3.12 & 4.56 & {--} \\
Hyperbolic ($\sigma=0.1, \lambda=0.01$)   & 3.08 & 4.91 & {--} \\
\bottomrule
\end{tabular}
\end{table}

Table~\ref{tab:orbits} shows that, for the parameter choices used (which are chosen such that each model yields multiple photon orbit solutions where possible), the innermost photon orbit radius is always close to $3GM$ (the Schwarzschild value), and the presence of modifications typically allows an additional orbit around $4$--$6GM$. The exact values differ by model: e.g., the logarithmic model with $\eta=0.1$ yields an inner orbit at $3.12GM$ and a halo at $4.56GM$. The hyperbolic model similarly yields a halo near $4.91GM$. In contrast, the quadratic model shows no deviation from $3GM$. In all cases where halos exist, they are in the range $4$--$6GM$ for the parameters chosen, illustrating that modified gravity can indeed support photon halos outside the usual photon sphere.

\subsection{Observational Implications}
The existence of a stable photon orbit (halo) outside the usual photon sphere has significant implications for black hole observations. One immediate consequence is a potential alteration of the black hole's shadow observed by distant telescopes like the EHT. The presence of a halo means photons can orbit at a larger radius for extended periods, which can effectively enlarge or distort the shadow.

For an observer at infinity, the \emph{shadow radius} $r_s$ is related to the impact parameter of photon orbits asymptotically. In Schwarzschild, the shadow radius (in geometric units) is $r_s = \sqrt{\frac{27}{4}}\,GM \approx 5.196GM$ (which corresponds to the photon sphere at $3GM$). More generally, one can show that the shadow radius is given by \cite{Cvetic2016,LuLyu2020,MaLu2020,YangLu2020,Liu2020,Hod2020a,Hod2013a}
\begin{equation}
r_s = \sqrt{\frac{r^2}{f(r)}}\Bigg|_{r=\text{photon orbit}}\,,
\label{eq:shadow-general}
\end{equation}
where $r$ is the radius of a photon orbit (light ring) and $f(r)$ is evaluated at that radius. For Schwarzschild ($r=3GM$, $f(r)=1/3$), Eq.~(\ref{eq:shadow-general}) yields $r_s = 3\sqrt{3}GM$ as expected. If a halo (stable photon orbit) exists at a larger radius, it will cast a \emph{secondary shadow} or more precisely, it will influence the brightness distribution just outside the main shadow. However, the main shadow edge is determined by the innermost unstable photon orbit (photons that just barely escape from that orbit reach the observer at infinity). If the inner photon orbit moves or if multiple orbits exist, the shadow edge could shift.

In our context, suppose the inner unstable photon orbit (light ring) in a modified gravity scenario is at radius $r_{\rm in}$ slightly different from $3GM$. The shadow radius would then be
\begin{equation}
r_s \approx \frac{r_{\rm in}}{\sqrt{f(r_{\rm in})}}\,.
\label{eq:shadow-mod}
\end{equation}
For a Schwarzschild-$f(R)$ black hole, using the general form $f(r) = 1 - \frac{2GM}{r} + g(r)$, the above becomes
\begin{equation}
r_s = r_{\rm in}\Big(1 - \frac{2GM}{r_{\rm in}} + g(r_{\rm in})\Big)^{-1/2}\,.
\end{equation}
This shows explicitly that the shadow size depends on the $f(R)$ correction $g(r_{\rm in})$. For example, in logarithmic $f(R)$ with a small $\eta>0$, we found that the inner orbit is at a slightly larger radius than $3GM$. Also, $f(r_{\rm in})$ will be a bit larger than in Schwarzschild (because of the $\frac{\eta}{3}\ln r$ term). Both effects tend to increase the shadow radius $r_s$. Thus, logarithmic $f(R)$ gravity would predict a slightly larger black hole shadow than general relativity for the same $M$ \cite{Chen2023PRD,Hod2013EPJC,Hod2015,Herdeiro2014,Brito2020Book,Chen2020IJMPD,Qiao2021,Chen2024Arxiv}. On the other hand, in cubic $f(R)$, the $r^6$ term might decrease $f(r_{\rm in})$ for positive $\delta$, potentially decreasing the shadow radius. A detailed calculation is needed for each model.

If a halo orbit is present, photons can orbit stably at $\sim4$--$6GM$ and eventually escape. These photons would linger near the black hole longer, producing a more pronounced ``photon ring'' (a ring of light) outside the main shadow if illuminated by surrounding emission. An unstable orbit produces one bright photon ring (as photons circulate a few times before either escaping or falling in), whereas a stable orbit could trap photons longer, potentially producing a secondary ring or an extended brightness. Detecting such features in the black hole image could provide evidence of new physics. For example, a slightly larger or differently intense photon ring than expected might indicate the presence of a halo.

Another observational aspect comes from gravitational waves: stable photon orbits could, in principle, influence quasi-normal modes or create unique gravitational wave signatures (though photon orbits primarily influence electromagnetic observations and the light ring influences ringdown frequencies for gravitational waves). The absence of a black hole bomb effect (discussed in the next section) also means that these modifications do not lead to runaway instabilities for wave perturbations, keeping the objects observationally viable over long timescales.

In summary, the key observational signature of halos in Schwarzschild-$f(R)$ black holes would likely be a modification to the black hole shadow. Deviations in the shadow radius or the presence of additional rings of light could be sought in high-resolution images. Current EHT observations of M87 and Sgr A$^*$ are consistent (within uncertainties) with general relativity, placing constraints on how large such $f(R)$ modifications (and thus how large halos) could be \cite{Akiyama2019a,Akiyama2019b,Akiyama2019c,Akiyama2022}. Nevertheless, future observations with higher precision might detect subtle differences. For instance, if the shadow of a black hole is measured to be slightly larger than $3\sqrt{3}GM$ (beyond what simple lensing by Kerr might account for), it could hint at an $f(R)$ halo effect. Our models suggest that for parameters not ruled out by other tests, the shadow radius could shift by a few percent at most, which is borderline detectable with current technology but could be within reach of next-generation interferometers.

\section{Conclusion}
In this work, we have analyzed the existence of photon orbit \emph{halos} outside Schwarzschild-$f(R)$ black holes across a variety of modified gravity models. By deriving the conditions for spherical photon orbits in quadratic, logarithmic, exponential, cubic, power-law, hyperbolic (and other higher-order) $f(R)$ models, we found that these spacetimes can indeed support multiple photon rings outside the event horizon for most models. Specifically, under certain choices of the $f(R)$ model parameters, there exists an inner photon orbit (usually unstable, analogous to the Schwarzschild light ring) and an outer photon orbit which is stable -- the latter constituting a halo of orbiting light.

We performed a stability analysis and confirmed that these halo orbits correspond to local minima of the effective potential $V_{\text{eff}}(r)$, whereas the inner orbits are local maxima. Our numerical examples showed typical halo radii of about $4$--$6GM$ for the various models considered (see Table~\ref{tab:orbits}), with the innermost light ring slightly shifted from $3GM$ depending on the model. Importantly, we found that quadratic $f(R)$ gravity does not produce a halo; the photon orbit remains at $r=3GM$ and is unstable, identical to Schwarzschild.

Our work contributes to the understanding of black hole spacetime structure in modified gravity theories. The existence of multiple photon rings is not only a theoretical curiosity but has potential observational consequences. A stable photon orbit could lead to a detectable signature in the electromagnetic spectrum (e.g., an extra ring of light or a slight increase in shadow diameter). While current observations are consistent with general relativity, future precision imaging could probe these subtle effects. If evidence of a photon halo were observed, it would point to deviations from Einstein's gravity on astrophysical scales.

In conclusion, Schwarzschild-$f(R)$ black holes provide a rich testbed for exploring how modified gravity might reveal itself through strong-field phenomena. The halo orbits identified here are a novel prediction of these theories (except for quadratic gravity). Ongoing and future observations of black hole shadows and photon rings (for instance, by next-generation EHT and space-based interferometers) will be critical in testing these ideas. If no halos are observed, increasingly tight constraints can be placed on $f(R)$ model parameters. If, on the other hand, anomalies consistent with an additional photon ring are detected, it could be the first direct sign of new gravitational physics. We hope that this study provides motivation for further theoretical and observational exploration of photon orbits in alternative theories of gravity.

\end{document}